\documentclass{article}

\usepackage{arxiv}
\usepackage{authblk}

\usepackage{hyperref} % For hyperlinks in the PDF
\usepackage{graphicx} % include graphics
\usepackage[nolist]{acronym}
\usepackage{amsmath}
\usepackage{bbm}
\usepackage{todonotes}
\usepackage{upgreek}
\usepackage{tikz}
\usetikzlibrary{decorations.pathreplacing,angles,quotes}

\usepackage[%
  backend=biber,
  url=false,
  style=alphabetic,
  maxnames=4,
  minnames=3,
  maxbibnames=99,
  giveninits,
  uniquename=init]{biblatex}

\bibliography{references}

\begin{document}

%\articletype{Paper} %	 e.g. Paper, Letter, Topical Review...

\title{A model for positron annihilation in multi-layer systems by solving the diffusion equation using different positron affinities}

%\author{Lucian Mathes$^{1,2}$, Michael G\"oldl$^{1,2}$, Michael Leitner$^1$,
%        Bettina Kohlhaas$^1$, Maximilian Suhr$^1$, Vassily Vadimovitch Burwitz$^{1,2,3}$,
%        Armin Manhard$^3$, Christoph Hugenschmidt$^1$}
%\affil{$^1$Heinz Maier-Leibnitz Zentrum (MLZ), Technical University of Munich, Lichtenbergstr. 1, 85748 Garching, Germany}
%\affil{$^2$School of Natural Sciences, Physics Department, Technical University of Munich, James-Franck-Str. 1, 85748 Garching, Germany}
%\affil{$^3$Max Planck Institute for Plasma Physics, Boltzmannstr. 2, 85748 Garching, Germany}
%\email{lucian.mathes@tum.de}

%%% FOR ARXIV PREPRINT %%%
\renewcommand\Authfont{\bfseries}
\setlength{\affilsep}{0em}
\author[1,2]{Lucian Mathes\thanks{\texttt{lucian.mathes@tum.de}}}
\author[1,2]{Michael G\"oldl}
\author[1]{Michael Leitner}
\author[1]{Bettina Kohlhaas}
\author[1,2]{Maximilian Suhr}
\author[1,2,3]{Vassily Vadimovitch Burwitz}
\author[3]{Armin Manhard}
\author[1]{Christoph Hugenschmidt}
\affil[1]{Heinz Maier-Leibnitz Zentrum (MLZ), Technical University of Munich, Lichtenbergstr. 1, 85748 Garching, Germany}
\affil[2]{School of Natural Sciences, Physics Department, Technical University of Munich, James-Franck-Str. 1, 85748 Garching, Germany}
\affil[3]{Max Planck Institute for Plasma Physics, Boltzmannstr. 2, 85748 Garching, Germany}

\renewcommand{\shorttitle}{A model for positron annihilation in multi-layer systems}
\maketitle

\begin{acronym}
    \acro{limpid}[LIMPID]{Layer-wise Investigation of Measurements on Positron Implantation and Diffusion}
    \acro{pas}[PAS]{Positron Annihilation Spectroscopy}
    \acro{dbs}[DBS]{Doppler-Broadening Spectroscopy}
    \acro{rbs}[RBS]{Rutherford Backscattering Spectrometry}
\end{acronym}

\begin{abstract}
%We present \acs{limpid}, an open-source Python-based computational tool that solves the positron diffusion equation in multilayer systems, incorporating material-specific implantation profiles, diffusion parameters, and positron affinities.
We present a method for solving the positron diffusion equation in multi-layer systems.
Our approach incorporates material-specific implantation profiles, diffusion parameters, and positron affinities.
It utilizes a Markov chain approach to model annihilation probabilities and provides fitting capabilities for experimental $S$ (lineshape) parameter data.
We have implemented this algorithm in Python and made it available for free under the name \acs{limpid}.
To demonstrate its performance, we analyze depth-resolved \acl{dbs} measurements of a Cu layer on a Si substrate, achieving excellent agreement with the experimental profiles.
The \acs{limpid} tool enhances the reproducibility and comparability of positron defect characterization measurements across different research groups.
\end{abstract}

\keywords{Positron annihilation \and Positron diffusion \and Doppler-broadening spectroscopy}

\section{Introduction}
% PAS
\ac{pas} is a well-established and highly sensitive technique used to study atomic-level defects in solid materials.
%It relies on the interaction of positrons with the material under investigation.
After implantation, positrons rapidly thermalize in the crystal lattice \cite{schultz1988}.
Once thermalized, they diffuse through the lattice with a diffusion length that strongly depends on the defect concentration \cite{puska1994}.
Finally, the positron annihilates with an electron by the emission of gamma radiation, in most cases consisting of two 511~keV photons.
The lifetime of positrons depends on the local electron density, while the energy of the annihilation radiation provides insights into the electron momentum distribution.
This information, in turn, reveals the nature of defects present in the material.
\ac{pas} is particularly sensitive to distinguishing between delocalized positrons annihilating in a defect-free lattice and positrons that are trapped in lattice imperfections such as vacancies, dislocations, and voids \cite{cizek2018}.

% DBS in general
A widely used method within \ac{pas} is \ac{dbs} of the positron-electron annihilation line.
The annihilating electrons transfer their momentum to the annihilation photons (the momentum of the thermalized positrons is negligible).
The electron momentum distribution, hence, leads to Doppler-shifted annihilation photons, which produce a Doppler-broadened annihilation photo peak in the recorded gamma spectrum.
The broadening of the annihilation peak, which differs between defect-free bulk and defects, is commonly quantified by the lineshape parameter, $S$, which represents the fraction of events within a defined low-momentum region around the center of the Doppler-broadened annihilation peak.
% depth-resolved DBS
By employing monoenergetic positron beams, depth-resolved \ac{dbs} measurements can be performed.
These allow the measurement of the $S$ parameter as a function of positron implantation energy, $S(E)$, and hence the investigation of, e.g., near-surface region defects, layered structures, and defect profiles.
The $S(E)$ profile contains information about the thickness of layers as well as the positron diffusion length in each layer and thus the defect concentration.
%the investigation to probe how positrons diffuse within the material and interact with defects at different depths.
%The $S(E)$ profile is shaped by and, therefore, contains information about the thickness of layers as well as the positron diffusion length and in turn of the defect concentraion.

% LIMPID
Due to the diffusion of thermalized positrons before annihilation, the annihilation profile, i.e., the number of positrons that annihilate in each layer and at the surface, differs from the implantation profile.
%Modeling only the positron implantation in a material is not sufficient to obtain the so-called annihilation profile, i.e., the number of positrons that annihilate in each layer and at the surface.
%We also have to take positron diffusion into account; this is where \ac{limpid} comes into play.
The diffusion behavior of positrons is influenced by various material properties, including defect concentration, layer structures, and positron affinity.
To accurately extract defect-related parameters from depth-resolved \ac{dbs} experiments, it is necessary to solve the positron diffusion equation, which describes how positrons migrate through the material before annihilation. 
Since complex boundary conditions and material-specific properties can influence diffusion, numerical modeling is often required to interpret experimental data effectively.
\ac{limpid} is a computational tool specifically developed to solve this diffusion problem efficiently across a wide range of systems and boundary conditions.
%It provides quantitative estimates of positron diffusion lengths, from which the vacancy concentration within a solid can be determined.
While the diffusion length, from which the vacancy concentration can be determined, is typically the primary output of \ac{limpid}, other material parameters -- such as layer thicknesses or positron affinities -- can also be fitted, provided that a sufficient number of known/fixed parameters is available.

%\subsection{The Need for a Software Tool}
Depth-resolved \ac{dbs} relies on the implantation of monoenergetic positrons into a sample, resulting in a broad, peak-shaped implantation profile. 
Since positrons diffuse within the material before annihilation, accurately modeling their diffusion behavior is crucial for interpreting data from depth-resolved \ac{dbs} experiments. 
The \ac{limpid} algorithm is specifically designed to solve the diffusion problem for thermalized positrons in solid materials.

% requirements
Starting from a given positron implantation profile as described by, e.g., Makhov \cite{asoka-kumar1990, makhov1960_2}, \ac{limpid} models the positron diffusion and returns an $S$ parameter value for each individual positron implantation energy.
We have implemented an optional correction for epithermal positrons at low positron implantation energies.
In the case of layered structures, \ac{limpid} accounts for multiple layers of varying thicknesses, as well as material-dependent diffusivities, lifetimes, $S$ parameter, and positron affinities.

To ensure accessibility and auditability of use for researchers across different scientific disciplines, \ac{limpid} was implemented in Python and published as open-source\footnote{\ac{limpid} is licensed under the \href{https://www.gnu.org/licenses/gpl-3.0.en.html}{GNU General Public License v3.0}. Source code available at \url{https://github.com/lucianmathes/limpid}.}.
Python was chosen due to its widespread adoption in the scientific community, the extensive ecosystem of numerical and data analysis libraries, and its user-friendly syntax.
The algorithm represents samples in a class-based digital structure, allowing users to input and access experimental parameters and sample properties as class attributes.
For parameter estimation and curve fitting, \ac{limpid} leverages the lmfit library \cite{lmfit}, which provides a framework for non-linear least-squares fitting.
The core numerical computations rely on NumPy \cite{numpy} and SciPy \cite{scipy}.
Matplotlib \cite{matplotlib} handles the visualization of results.

\ac{limpid} was developed to support positron research groups in data analysis, aiming to replace both an earlier tool introduced by van Veen \cite{veen1991} and the many custom-built and unpublished scripts tailored to specific experimental setups and sample structures.
By providing a unified, improved, expandable, and open-source solution, \ac{limpid} enhances the transparency, reproducibility, and comparability of depth-resolved \ac{dbs} analyses across research groups.
The algorithm has already been successfully applied, as demonstrated in recent publications \cite{schalk2024, burwitz2025}.

%------------------------------------------------

\section{Method / Algorithm}

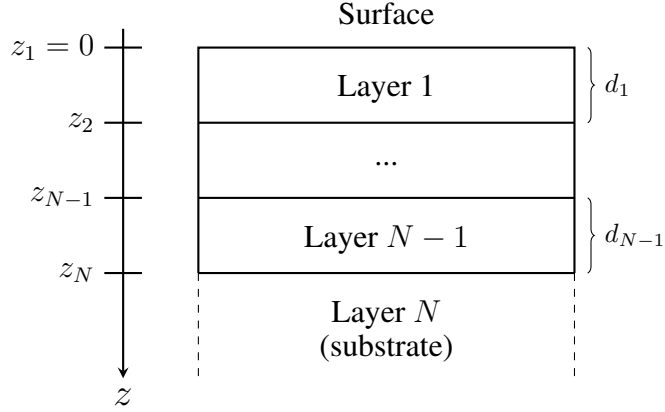
\begin{figure}[!ht]
    \centering
    \begin{tikzpicture}
        \draw [thick] (0,0) -- (0,3) -- (5,3) -- (5,0);
        \draw [thick] (0,1) -- (5,1);
        \draw [thick] (0,2) -- (5,2);
        \draw [thick] (0,0) -- (5,0);
        \draw [dashed] (0,0) -- (0,-1.4);
        \draw [dashed] (5,0) -- (5,-1.4);

        \draw [-stealth, thick](-1, 3.25) -- (-1, -1.4) node [below] {\Large $z$};
        \draw [thick](-1.25, 3) node [left] {\large $z_1=0$} -- (-0.75, 3);
        \draw [thick](-1.25, 2) node [left] {\large $z_2$} -- (-0.75, 2);
        \draw [thick](-1.25, 1) node [left] {\large $z_{N-1}$} -- (-0.75, 1);
        \draw [thick](-1.25, 0) node [left] {\large $z_{N}$} -- (-0.75, 0);

        \node at (2.5, 3.45) {\large Surface};
        \node at (2.5, 2.45) {\large Layer 1};
        \node at (2.5, 1.45) {\large ...};
        \node at (2.5, 0.45) {\large Layer $N-1$};
        \node at (2.5, -0.55) {\large Layer $N$};
        \node at (2.5, -1) {\large (substrate)};

        \draw [decorate, decoration={brace,raise=5pt}] (5,3) -- node [right=8pt] {$d_1$} (5,2);
        \draw [decorate, decoration={brace,raise=5pt}] (5,1) -- node [right=8pt] {$d_{N-1}$} (5,0);

    \end{tikzpicture}
    \caption{
        Schematic drawing of a generic sample explaining the nomenclature used by \ac{limpid}.
        The sample consists of $N$ layers, whereby the $N^{\textrm{th}}$ layer has an infinite thickness.    }
    \label{fig:sample}
\end{figure}

Positrons with a given kinetic energy interact with the sample material, losing energy through inelastic collisions. 
This process, known as thermalization, continues until the positrons reach thermal equilibrium with their surroundings, at an energy of approximately 40~meV at room temperature.

\begin{enumerate}
    \item \textbf{Positron implantation:}
        Initially, positrons are implanted into the material.
        The depth distribution is calculated using an implantation model (here: Makhov) and a set of material-dependent parameters.
\end{enumerate}

\noindent
The transport of thermalized positrons in a solid is modeled using diffusion theory.
%In \ac{limpid}, the diffusion process for a given positron implantation energy is divided into two steps:
In \ac{limpid}, the sample is modelled via a number of homogeneous layers of arbitrary thickness and positron-relevant properties.
The trajectory of a given positron is conceptually divided into segments corresponding to its residing in the individual sample layers.
The resulting mathematical problem is thus solved in two stages:

\begin{enumerate}
    \setcounter{enumi}{1}
    \item \textbf{Diffusion to the layer boundaries:}
        After thermalization, positrons diffuse  through the material via random thermal motion, eventually either annihilating within a layer or reaching an interface between layers.
        In a homogeneous sample, this process simplifies to positron diffusion to the surface or annihilation in the bulk.
    \item \textbf{Jumping between layer boundaries:}
        For those positrons that reach a boundary before annihilating, the subsequent trajectory is modeled as a Markov process.
        It is defined by the probabilities of annihilating within one of the two adjacent layers or reaching one of the two neighbouring boundaries, tracking how positrons move through the material until they eventually disappear.
        %If positrons reach a boundary, they may either transition to an adjacent layer boundary or annihilate during the transition.
        %This step is modeled as a Markov process, tracking how positrons move through the material until they eventually disappear.
\end{enumerate}

\noindent
Both steps include positron annihilation at various depths within the material, resulting in an annihilation profile that reflects the contribution of positrons from each layer or the surface to the measured signal.
From this profile, \ac{limpid} calculates a weighted $S$ parameter from the layer- and surface-specific $S$ parameters.
Figure~\ref{fig:sample} shows a schematic drawing of a generic sample clarifying the nomenclature and indexing used in the following.

\subsection{Positron Implantation}
Positron implantation and thermalization are complex and consist of many particle-particle interactions.
The distribution of thermalized positrons, i.e., the implantation profile, at a given energy $E$ is already broad and can be approximated by its width $\propto E^{1.62}$ \cite{vehanen1987}.
%The distribution becomes even broader through diffusion.
%To accurately model the implantation profile, time-consuming Monte Carlo simulations can be performed.
%For a simpler model and everyday use, these results can be fitted using an appropriate function as determined by Makhov \cite{asoka-kumar1990, makhov1960_2} or Ghosh \cite{ghosh1995}.
Typically, the Makhov distribution is employed to model the implantation profile \cite{asoka-kumar1990, makhov1960_2}, although more sophisticated models also exist \cite{ghosh1995}.
Parameters for a variety of materials have been determined through Monte Carlo simulations and are publicly available \cite{dryzek2008, puska1994}.
Multi-layer implantation profiles are combined from the individual material-specific implantation profiles scaled with the correct fractions \cite{aers1994}.
\ac{limpid} currently includes the Makhov implantation profile described by:
\begin{equation}
    \label{eq:makhov}
    P(z) = \frac{m z^{m-1}}{z_0^m} \, \exp\bigg[ -\bigg(\frac{z}{z_0}\bigg)^m \bigg],
\end{equation}
with
\begin{equation}
    \label{eq:makhov_z0}
    z_0 = \frac{A}{\rho \, \Gamma (1 + \frac{1}{m})} \, E^n,
\end{equation}
where $m$, $n$ and $A$ are material-specific parameters, $\rho$ is the material density and $E$ is the positron implantation energy.
Two example profiles for a layered system are shown in Figure~\ref{fig:cusi-implantation}.
% in outlook:  but we are planning to include more in the future.

\subsection{Diffusion to the Layer Boundaries}

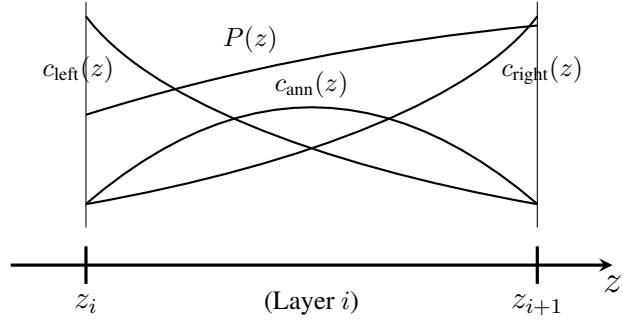
\begin{figure}[!ht]
    \centering
    \begin{tikzpicture}
        \draw [-stealth, very thick](-4, 0.5) -- (4, 0.5) node [below] {\Large $z$};
        \draw [very thick](-3, 0.25) node [below] {\large $z_{i}$} -- (-3, 0.75);
        \draw [very thick](3, 0.25) node [below] {\large $z_{i+1}$} -- (3, 0.75);

        \draw (-3, 1) node [right] {} -- (-3, 4) node [above] {};
        \draw (3, 1) node [right] {} -- (3, 4) node [above] {};

        %\draw [dashed] (0, 4) -- (0, 0.15) node [below] {$z_0$};
        %\draw (-0.25, 1) -- (0.25, 1) node [right] {\Large $a$};

        \node at (0,0) {(Layer $i$)};

        \draw [thick] (-3,2.5) arc(120:100:18.4 and 10);
        \draw (-1.3, 3.5) node [right] {$P(z)$};

        \draw [thick] (3,3.81) arc(-13:-60:12.66 and 3.9);
        \draw [thick] (-3,3.81) arc(-167:-120:12.66 and 3.9);
        \draw[thick, domain =-3:3, variable = \x] plot({\x*1},{-\x*\x/7 + 2.6});

        \draw (-3.7, 3.1) node [right] {$c_{\textrm{left}}(z)$};
        \draw (2.4, 3.1) node [right] {$c_{\textrm{right}}(z)$};
        \node at (0, 2.9) {$c_{\textrm{ann}}(z)$};
    \end{tikzpicture}
    \caption{
    Schematic drawing of the first diffusion step inside the $i^{th}$ layer.
    $P(z)$ represents the positron implantation profile within a material layer of thickness $d$.
    %After diffusion, the positron distribution at the layer boundaries is determined by integrating the product of $P(z)$ and the diffusion rates $c_{\textrm{left}}(z)$ and $c_{\textrm{right}}(z)$.
    After diffusion, the positron distribution at the layer boundaries is determined by integrating the product of $P(z)$ and the probability of a positron reaching the left [right] boundary $c_{\textrm{left}}(z)$ [$c_{\textrm{right}}(z)$].
    The corresponding mathematical formulation is provided in Equation~\ref{eq:n_left}.
    $c_{\textrm{ann}}$ is the probability of a positron to annihilate before reaching a boundary.
    All three probabilities sum to 1.
    }
    \label{fig:diffusion_to_boundaries}
\end{figure}

The one-dimensional diffusion equation inside layer $i$ is given by
\begin{equation}
    \label{eq:diff1}
    \frac{\partial n(z, t)}{\partial t} = D_i \frac{\partial^2 n(z,t)}{\partial z^2} - \lambda_{\textrm{eff},i} \, n(z, t),
\end{equation}
where $n(z, t)$ is the positron density as a function of time $t$ and depth $z$, $D_i$ is the diffusion coefficient, and $\lambda_{\textrm{eff},i}$ is the effective annihilation rate,
\begin{equation}
    \lambda_{\textrm{eff},i} = \frac{1}{\tau_{\textrm{eff},i}} = \lambda_{\textrm{bulk},i} + \kappa c_i,
\end{equation}
where $\lambda_{\textrm{bulk},i}$ is the annihilation rate of positrons in the defect-free bulk, $\kappa$ is the specific trapping rate of a defect, and $c_i$ is the defect concentration inside layer $i$.
In the case of multiple types of defects, $j$, present in the layer, $\kappa c_i$ becomes $\sum_j \kappa_j c_{j,i}$ representing an ``effective'' defect with ``effective'' $S$ parameter.

Under steady-state conditions, we get
\begin{equation}
    \label{eq:diff2}
    0 = D_i \, \frac{\textrm{d}^2 n(z)}{\textrm{d} z^2} - \lambda_{\textrm{eff},i} \, n(z) + \mu P(z,E),
\end{equation}
where $P(z,E)$ is the implantation profile for energy $E$ and $\mu=1~s^{-1}$ is the positron flux.
Note that all positron fractions calculated in the following are independent of the value for $\mu$.
Solving Equation~\ref{eq:diff2} gives the fraction of positrons diffusing to the left (low $z$) layer boundary instead of directly annihilating or reaching first the right boundary:
\begin{equation}
    \label{eq:n_left}
    f_{\textrm{left},i} =  \int_{z_i}^{z_{i+1}} P(z,E) \, c_{\textrm{left},i}(z) \, \textrm{d}z
                        = \int_{z_i}^{z_{i+1}} P(z,E) \, \frac{\sinh[u_i(z_{i+1}-z)]}{\sinh[u_i(z_{i+1}-z_i)]} \, \textrm{d}z,
\end{equation}
where $c_{\textrm{left},i}(z)$ is the probability of a positron at position $z$ diffusing to the left boundary (as plotted in the explanatory graphic in Figure~\ref{fig:diffusion_to_boundaries}) and $z_{i+1}-z_i$ is equal to the layer thickness; 
$u_i$ is the inverse positron diffusion length inside layer $i$:
\begin{equation}
    u_i = \sqrt{\frac{\lambda_{\textrm{eff},i}}{D_i}} = \frac{1}{L_{i}}.
\end{equation}
We get the fraction of positrons diffusing to the right (high $z$) layer boundary, $f_{\textrm{right},i}$, by replacing $c_{\textrm{left},i}(z)$ with
\begin{equation}
    \label{eq:n_right}
    c_{\textrm{right},i}(z) = \frac{\sinh[u_i(z-z_i)]}{\sinh[u_i(z_{i+1}-z_i)]}.
\end{equation}
%Both integrals are visualized in Figure~\ref{fig:diffusion_to_boundaries}.
The fraction of positrons annihilating inside the $i^{th}$ layer is then
\begin{equation}
    f_{\textrm{ann},i} = 1 - f_{\textrm{left},i} - f_{\textrm{right},i}.
\end{equation}

\subsection{Diffusion Between Layer Boundaries}\label{sec:diff2}

\begin{figure}[!ht]
    \centering
    \begin{tikzpicture}
        \node at (2.3,0) {(Layer $i$)};
        \draw [-stealth, very thick](-0.5, 0.5) -- (5.5, 0.5) node [below] {\Large $z$};

        \draw [very thick](0, 0.25) node [below] {\large $z_i$} -- (0, 0.75);
        \draw [very thick](4.5, 0.25) node [below] {\large $z_{i+1}$} -- (4.5, 0.75);

        \draw (0, 1) node [right] {} -- (0, 4);
        \draw (4.5, 1) node [right] {} -- (4.5, 4);

        \draw [-stealth, thick](0, 3.5) -- (2, 3.5) node [above, midway] {\Large $N_{\textrm{right},i}$};
        \draw [-stealth, thick](2, 3.5) -- (4.5, 3.5) node [above, midway] {\Large $J_{\textrm{right},i}$};

        \draw [-stealth, thick] (2, 3.5) arc(90:0:0.25 and 2.5);
        \draw (2, 1) -- (2.5, 1) node [right] {\Large $A_{\small\textrm{right},i}$};
    \end{tikzpicture}
    \caption{
    Schematic drawing of the fluxes in the second diffusion step.
    After reaching a layer boundary, positrons either jump between boundaries or annihilate within the layer.
    The transition probabilities calculated from the fluxes form a Markov chain, which describes how positrons move through the system until they are annihilated.
    }
    \label{fig:diffusion_between_boundaries}
\end{figure}
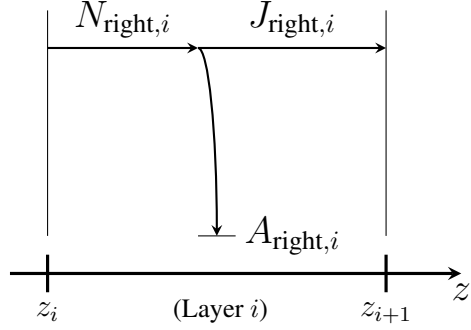

After the initial diffusion step, which is essentially the set-up for the second/main diffusion step, the diffusion equation simplifies.
Now, positrons that reach a boundary can either transition to an adjacent boundary or annihilate.
Note that positrons can only annihilate inside layers or at the surface, but not at the infinitesimally thin layer boundaries, which are assumed not to trap positrons.
(Interfaces with high positron affinity/attractiveness trapping positrons -- e.g., due to lattice mismatch -- can be modelled as additional thin layers.)
This process follows
\begin{equation}
    \label{eq:diff3}
    0 = D_i \frac{\textrm{d}^2 n(z)}{\textrm{d} z^2} - \lambda_{\textrm{eff},i} \, n(z),
\end{equation}
where $D_i$ is the diffusion coefficient and $\lambda_{\textrm{eff},i}$ is the effective annihilation rate for the material of layer $i$.
We define the positron density at the layer boundary $z_i$ as
\begin{equation}
    \label{eq:x}
    n(z_i) = n_0
\end{equation}
with units positrons/m.
The solution of the diffusion equation is then
\begin{equation}
    \label{eq:diff-sol}
    n(z) = n_0\,\frac{\sinh[u_i(z-i_{i+1})]}{\sinh[u_i(z_i-i_{i+1})]}.
\end{equation}
We get the flux of positrons leaving the boundary towards the right (with units positrons/s) using Fick's first law of diffusion and the right-sided derivative:
\begin{equation}
    \label{eq:diff5}
    N_{\textrm{right},i} = -D_i \left. \frac{\mathrm{d}n(z)}{\mathrm{d}z} \right|_{z=z_i^+} = n_0 \, u_i D_i \, \frac{\cosh[u_i(z_{i+1}-z_i)]}{\sinh[u_i(z_{i+1}-z_i)]}.
\end{equation}
The flux of positrons that reach the next (right-sided) layer boundary is given by the left-sided derivative at $z_{i+1}$:
\begin{equation}
    \label{eq:diff4}
    J_{\textrm{right},i} = -D_i \left. \frac{\mathrm{d}n(z)}{\mathrm{d}z} \right|_{z=z^-_{i+1}}
    = \frac{n_0 \, u_i D_i}{\sinh[u_i(z_{i+1}-z_i)]}.
\end{equation}
The amount of positrons annihilating on the way results from the difference between the two fluxes:
\begin{equation}
    \label{eq:ann}
    A_{\textrm{right},i} = N_{\textrm{right},i} - J_{\textrm{right},i}.
\end{equation}

Same applies to the left-sided diffusion and annihilation, where we obtain the fluxes $N_{\textrm{left},i}$, $J_{\textrm{left},i}$, and $A_{\textrm{left},i}$ by exploiting the symmetry of the diffusion process:
\begin{equation}
    \label{eq:diff7}
    A_{\textrm{left},i} = A_{\textrm{right},i-1},
\end{equation}
where $i$ is the layer boundary index.
Same applies to $J_{\textrm{right/left}}$ and $N_{\textrm{right/left}}$.
We normalize those fluxes ($n_0=1$) to get the probabilities of a positron diffusing to the next boundary and a positron annihilating on the way:
\begin{equation}
    \label{eq:ann}
    j_{\textrm{right},i} = \frac{J_{\textrm{right},i}}{N_{\textrm{right},i} + N_{\textrm{left},i}},
\end{equation}
(equivalently for $j_{\textrm{left}}$, $a_{\textrm{right}}$, and $a_{\textrm{left}}$).
Note that the resulting probabilities depend on the diffusion and annihilation characteristics of both adjacent layers ($u_{i-1} D_{i-1}$ and $u_i D_i$).

We use a time-homogeneous Markov chain with a finite state space to describe the diffusion between layer boundaries.
The layer boundaries (plus the surface) are transient states, while the layers themselves (plus the surface) are absorbing states.
We implement this Markov process using matrices.
The matrix, $\mathbf{Q}$, is filled with the probabilities of transitions between transient states, i.e., positrons jumping from one layer boundary to another (including the surface).
For example, with 4 layers and therefore 3 layer boundaries (plus the surface), we get
\begin{equation}
    \mathbf{Q} = 
\begin{pmatrix}
    0 & j_{\textrm{left},1} & 0 & 0 \\
    j_{\textrm{right},0} & 0 & j_{\textrm{left},2} & 0 \\
    0 & j_{\textrm{right},1} & 0 & j_{\textrm{left},3} \\
    0 & 0 & j_{\textrm{right},2} & 0 \\
\end{pmatrix},
\end{equation}
where the $j_{\textrm{left/right},i}$ is the probability of a positron on layer boundary $i$ to jump to the left or right adjacent layer boundary (with 0 being the surface and $j_{\textrm{right},0}=0$).
Note that the surface here is both a boundary (in the sense that positrons can diffuse there) and a layer (in the sense that positrons can annihilate there), and is assumed to be infinitely attractive.
The matrix, $\mathbf{R}$, is filled with the probabilities of transitions from transient to absorbing states, i.e., positrons annihilating in a layer (including the surface).
For the above example:
\begin{equation}
    \mathbf{R} = 
\begin{pmatrix}
    1 & 0 & 0 & 0 \\
    0 & a_{\textrm{left},1} & 0 & 0 \\
    0 & a_{\textrm{right},1} & a_{\textrm{left},2} & 0 \\
    0 & 0 & a_{\textrm{right},2} & a_{\textrm{left},3} \\
    0 & 0 & 0 & a_{\textrm{right},3} \\
\end{pmatrix},
\end{equation}
where $a_{\textrm{left/right},i}$ is the probability of a positron on layer boundary $i$ to annihilate in the left or right layer of the boundary.
The matrix of absorption probabilities can be calculated as
\begin{equation}
    \label{eq:matrix3}
    \mathbf{B} = \mathbf{R} \cdot \sum_{n=0}^{\infty} \mathbf{Q}^n = \mathbf{R} \cdot (\mathbbm{1}-\mathbf{Q})^{-1},
\end{equation}
where we utilize the geometric series, and the resulting annihilation fractions per layer (and surface) for a given energy $E$ are
\begin{equation}
    \label{eq:matrix4}
    \vec{n}_{\textrm{ann}} = \mathbf{B} \cdot \vec{n} + \vec{f}_{\textrm{ann}},
\end{equation}
with the remaining positrons on boundaries, $\vec{n}$, and the already annihilated positrons, $\vec{f}_{\textrm{ann}}$, from the first diffusion step.
%Therefore, its dimension is equal to the number of layers plus the surface.
%To get the distribution of positrons after six steps, which is definitely sufficient for all positrons to annihilate, we multiply the distribution vector after the first diffusion step, $\vec{n}$, with the 6$^{th}$ power of the matrix:
%\begin{equation}
%    \label{eq:diff7}
%    \vec{n}_{\textrm{ann}} = \vec{n} \cdot \mathbf{M}^6.
%\end{equation}
%The result, $\vec{n}_{\textrm{ann}}$ is the annihilation distribution of the positrons in all layers and at the surface for a given implantation energy $E$.
The lineshape parameter $S$ results from a weighted sum of all layer- and surface-specific lineshape parameters
\begin{equation}
    \label{eq:sparam1}
    S = \vec{n}_{\textrm{ann}} \cdot \vec{S} = \sum_{i=1}^N S_i n_i + S_{\textrm{surf}} n_{\textrm{surf}},
\end{equation}
where $N$ is the number of layers.

\subsection{Depth Profiles}
Modelling diffusion and calculating annihilation profiles for multiple positron implantation energies yields a so-called depth profile:
\begin{equation}
    \label{eq:sparam2}
    S(E) = \sum_{i=1}^N S_i n_i(E) + S_{\textrm{surf}} n_{\textrm{surf}}(E).
\end{equation}
Fitting this model function to data obtained from depth-resolved \ac{dbs} experiments allows the extraction of the positron diffusion length and other parameters.

\subsection{Positron Affinity}
The positron affinity, $A_+$, quantifies how strongly positrons are attracted or repelled by different materials.
Values of the element dependent $A_+$ can be calculated from the chemical potential of electrons and positrons in the material and can be found in literature for a selection of materials \cite{puska1989}.
The positron affinity affects the mobility of positrons in layered systems and alloys with precipitates.
To account for the positron affinity, all diffusion and annihilation probabilities in the second step of modelling the diffusion ($a_{\textrm{left/right}}$, $j_{\textrm{left/right}}$, see section~\ref{sec:diff2}) are multiplied by a Boltzmann factor,
\begin{equation}
    \label{eq:affinity}
    \exp\bigg[-\frac{A_+}{k_{\tiny\textrm{B}}T}\bigg],
\end{equation}
where $k_{\textrm{B}}$ is the Boltzmann constant and $T$ the temperature, and subsequently normalized.

\subsection{Epithermal Correction}
A fraction of positrons may annihilate before complete thermalization with epithermal energy.
\ac{limpid} contains an optional simple correction to account for epithermal positrons as suggested by \cite{britton1988} and previously used by \cite{veen1991}.
It introduces an additional lineshape parameter value, $S_{\textrm{epi}}$, and an average epithermal scattering length, $L_{\textrm{epi}}$.
The scattering length $L_{\textrm{epi}}$ is typically in the order of 1~nm and, like any other variable, can either be fixed or varied during the fitting procedure.
The resulting $S(E)$ profile, including the epithermal correction, is given by
\begin{equation}
    \label{eq:sparam3}
    S(E) = S_{\textrm{epi}}\eta_{\textrm{epi}}(E) + (1-\eta_{\textrm{epi}}(E)) \Bigg[\sum_{i=1}^N S_i n_i(E) + S_{\textrm{surf}} n_{\textrm{surf}}(E)\Bigg],
\end{equation}
with the fraction of epithermal positrons,
\begin{equation}
    \eta_{\textrm{epi}} = \int_0^{\infty} P(z,E) \, \exp\bigg[- \frac{z}{L_{\textrm{epi}}}\bigg] \, \textrm{d}z.
\end{equation}

\section{Results / Example Application}
In this example, we use \ac{limpid} to analyze \ac{dbs} data obtained from measuring a Cu layer deposited on a Si substrate \cite{kohlhaas2022-ba}.
The Cu layer has been electron-beam physical vapor-deposited on B-doped Si with dimensions $10\times10\times0.4~\textrm{mm}^3$ with a deposition rate of $\approx0.1~\textrm{nm}/\textrm{s}$.
We calculate the positron implantation and annihilation fractions inside the layer structure.
%The Cr layer is 350~nm thick and has a positron diffusion length of 30~nm.
%The Si substrate has a positron diffusion length of 165~nm and is assumed to be infinitely thick.
Implantation profiles are calculated using the Makhov function (Equation~\ref{eq:makhov}) with the material-specific parameters listed in Table~\ref{tab:makhov-param}.
%Figure~\ref{fig:implantation-profile} shows two exemplary profiles for the implantation energies 10 and 15~keV.
The S parameter value for the Si substrate, as well as the positron diffusion length in Si, are known from \ac{limpid} fits to measurement data of a bare Si substrate.
% and Cr and Cu single layers on a Si substrate.
They are listed in Table \ref{tab:fit-params} and fixed for the fit.
The epithermal scattering length is fixed to $L_{\textrm{epi}}=1$~nm.

\bgroup
\def\arraystretch{1.1}%  1 is the default, change whatever you need
\begin{table}[!ht]
    \centering
    \caption{
    Material-specific Makhov parameters and positron affinities for Cu and Si taken from literature \cite{dryzek2008, puska1989}.
    %The Makhov parameters for Cr are calculated as the average of the parameters for Ti and Fe.
    }
    \label{tab:makhov-param}
    \begin{tabular}{ | c | c | c | c | c | c | }
      \hline
      & $\rho$ / g cm$^{-3}$ & $A$ / $\upmu$g cm$^{-2}$ keV$^{-n}$ & $m$ & $n$ & $A_+$ / eV \\
      \hline
      %Cr & 7.15 & 2.74 & 1.76 & 1.674 & $-2.62$ \\
      %\hline
      Cu & 8.96 & 2.84 & 1.73 & 1.67 & $-4.81$ \\
      \hline
      Si & 2.33 & 2.48 & 1.99 & 1.73 & $-6.95$ \\
      \hline
    \end{tabular}
\end{table}
\egroup

\bgroup
\def\arraystretch{1.3}%  1 is the default, change whatever you need
\begin{table}[!ht]
    \centering
    \caption{
    A list of fixed and varied \ac{limpid} fit parameters used for the Cu/Si system.
    The best-fit values on the left include the correct positron affinities and an epithermal correction.
    Note that the ``values w/o affinities'' correspond to a direct fit with no positron affinities provided and therefore do not relate to the red curve in Figure~\ref{fig:cusi} or the last fraction plot in Figure~\ref{fig:cusi-fractions}.
    }
    \label{tab:fit-params}
    \begin{tabular}{ | c | c | c || c | c | }
      \hline
        & Value & Type & Value w/o epithermal corr. & Value w/o affinities\\
      \hline
        $S_{\small\textrm{epi}}$ & $0.6308\pm0.0005$ & varied & -- & $0.6309\pm0.0006$ \\
      \hline
        $S_{\small\textrm{surf}}$ & $0.6208\pm0.0006$ & varied & $0.6269\pm0.0005$ & $0.6205\pm0.0006$ \\
      \hline
        $S_{\small\textrm{Cu}}$ & $0.5786\pm0.0004$ & varied & $0.5801\pm0.0006$ & $0.5775\pm0.0005$ \\
      \hline
        $S_{\small\textrm{Si}}$ & 0.6659 & fixed & 0.6659 & 0.6659 \\
      \hline
        $L_{\small\textrm{epi}}$ & 1~nm  & fixed & -- & 1~nm \\
      \hline
        $L_{\small\textrm{Cu}}$ & $(30.4\pm1.2)$~nm & varied & $(23.2\pm0.9)$~nm & $(32.1\pm1.3)$~nm \\
      \hline
        $L_{\small\textrm{Si}}$ & 386~nm & fixed & 386~nm & 386~nm \\
      \hline
        $d_{\small\textrm{Cu}}$ & $(448\pm3)$~nm & varied & $(448\pm5)$~nm & $(330\pm3)$~nm \\
      \hline
    \end{tabular}
\end{table}
\egroup

\begin{figure}[!ht]
    \centering
    \includegraphics[width=0.75\linewidth]{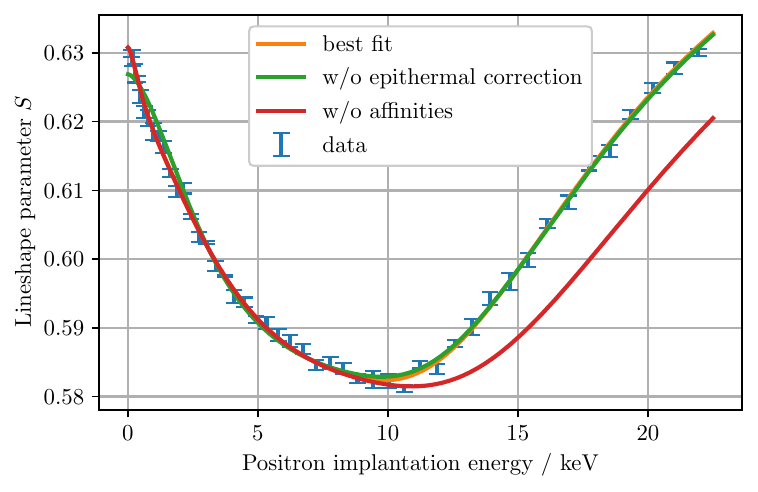}
    \caption{
        Depth-resolved \ac{dbs} data for a Cu layer on a Si substrate, fitted using the \ac{limpid} algorithm.
        The experimental data (symbols) and \ac{limpid} models (solid lines) demonstrate almost perfect agreement for the $S(E)$ profile.
        The orange line represents the best fit achieved (and is identical for the fit without positron affinities) using the correct positron affinities and an epithermal correction.
        The green line shows a fit with the epithermal correction disabled, resulting in significant deviation for energies $\leq3$~keV.
        The red line shows an affinity-less model of the best-fit result.
        All fit parameters are listed in Table~\ref{tab:fit-params}.
    }
    \label{fig:cusi}
\end{figure}

% Fit results
% Table
The free fit parameters and their best values are listed in Table~\ref{tab:fit-params} as type ``varied''.
% Interpretation
The Cu diffusion length, $L_{\textrm{Cu}}=30.4$~nm, appears to be short but reasonable for a vapor-deposited, presumably unordered and defect-rich layer.
The 448~nm layer thickness is close to what we aimed for with the vapor deposition duration ($\approx500$~nm).
Comparing the fit with and without positron affinities (see Table~\ref{tab:fit-params}) shows the value of the \ac{limpid} feature.
Both variations result in a nearly identical fit (see the ``best fit'' line in Figure~\ref{fig:cusi}), but the fit without positron affinities, i.e., assuming equal values for all materials, results in a slightly larger value for the positron diffusion length in Cu, $L_{\textrm{Cu}}=23.2$~nm, and a significantly smaller layer thickness, $d_{\textrm{Cu}}=330$~nm.
The fit with epithermal correction disabled yields similar results for diffusion length and layer thickness, but ceases to match the data towards low implantation energies $\leq3$~keV, as can be seen in Figure~\ref{fig:cusi}:
% Figure
It shows the measured $S$ parameter as a function of the positron implantation energy for the Cu/Si system, along with the corresponding \ac{limpid} fit.
Note that we did not include the fit without affinities, for it is identical to the best fit.
Instead, we include an affinity-less model of the best-fit results (see the red line in Figure~\ref{fig:cusi}) to illustrate the impact of the feature.
Modeling positron diffusion using the best fit results, but with the positron affinities set equal, $A_+^{\textrm{Si}}=A_+^{\textrm{Cu}}$, results in this significant deviation towards higher energies.
This can be attributed to the fraction of positrons diffusing from the (more attractive) Si towards Cu being reduced due to their different affinities compared to the affinity-less case.
%The resulting layer thicknesses are also small compared to \ac{rbs} results ($d_{\textrm{Cr}}=(103\pm1)$~nm and $d_{\textrm{Cu}}=(82\pm2)$~nm \cite{kohlhaas2022-ba}).
%The discrepancy between fitted thicknesses and \ac{rbs} measurements suggests that the deposited layers may be less dense or contain higher defect concentrations than bulk reference materials.
%This highlights the importance of considering layer density and microstructure when interpreting diffusion lengths and thicknesses.
%Fixing the layer thicknesses to the values determined by \ac{rbs} results in a slightly different fit (see green line in Figure~\ref{fig:cusi}).
%The parameters are still reasonable (even though the positron diffusion length in Cu is really short, $L_{\textrm{Cu}}=3~$nm) and are also listed in Table~\ref{tab:fit-params}.

\begin{figure}[!ht]
    \centering
    \includegraphics[width=0.75\linewidth]{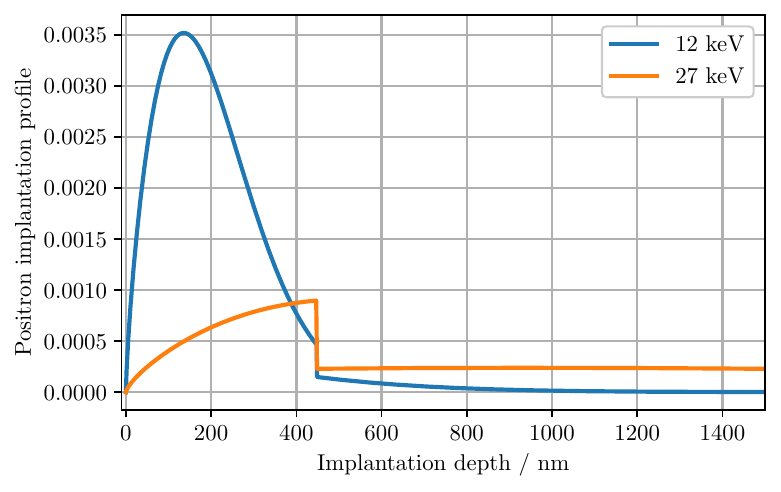}
    \caption{
    Calculated positron implantation profiles in the Cu/Si sample for two representative positron implantation energies (12~keV and 27~keV).
    We used the thicknesses resulting from the best fit in Figure~\ref{fig:cusi} and Table~\ref{tab:fit-params}, i.e., 448~nm Cu on top of the Si substrate.
    The profiles show how the implantation depth and spread depend on energy and (mainly) how density shapes multi-layer implantation.
    }
    \label{fig:cusi-implantation}
\end{figure}

As a result of the fit, we can use \ac{limpid} to visualize the multi-layer implantation profiles and the fractions of positrons annihilating in each channel.
Figure~\ref{fig:cusi-implantation} shows two exemplary implantation profiles for the positron implantation energies 12 and 27~keV.
The distinct profiles for both materials are calculated separately and concatenated in a way that assures the correct fractions in each layer and that they sum to 1.
As a result, we get jump discontinuities at the layer boundaries.

\begin{figure}[!ht]
    \centering
    \includegraphics[width=0.75\linewidth]{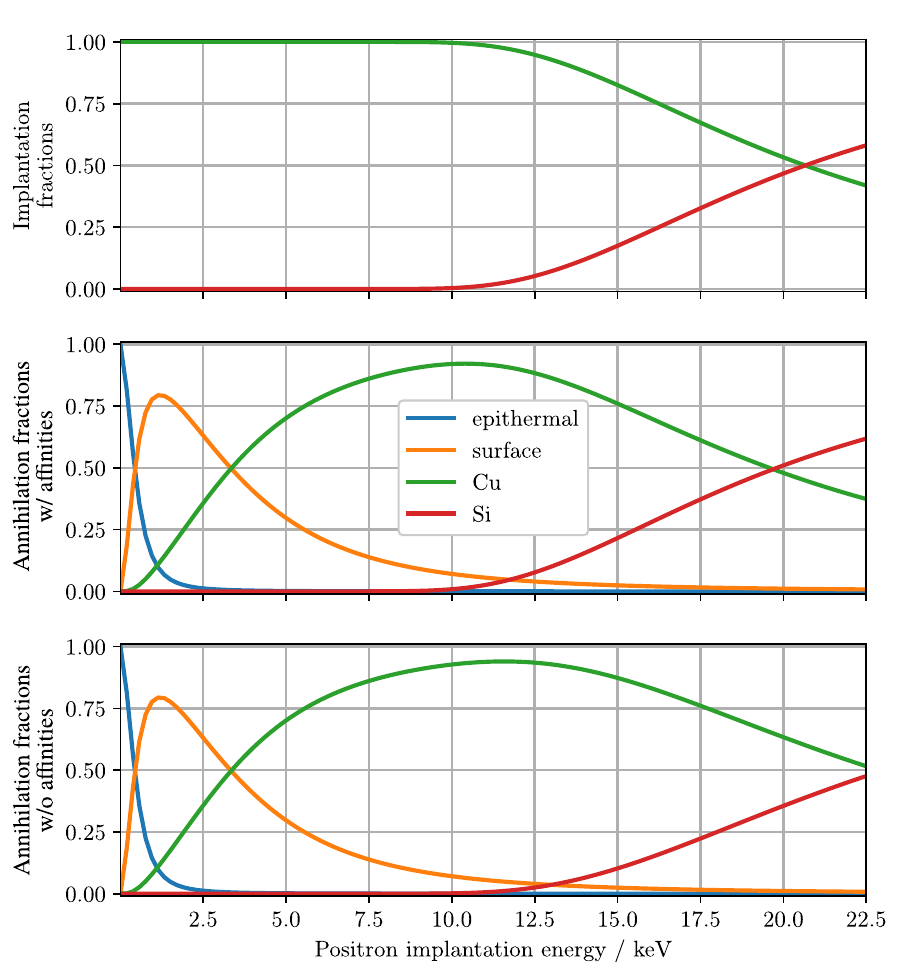}
    \caption{
    Fractions of positrons implanted and annihilated in different channels as a function of implantation energy.
    Top: Implantation fractions in each layer.
    Center: Corresponding annihilation fractions for the best fit results resolved into epithermal annihilation, surface annihilation, and annihilation in Cu and Si.
    Bottom: Corresponding annihilation fractions for an affinity-less model of the best-fit results.
    }
    \label{fig:cusi-fractions}
\end{figure}

Figure~\ref{fig:cusi-fractions} contains fraction plots of the implanted positrons and the different annihilation channels.
It aids in understanding the impact of different parameters, such as densities, layer thicknesses, and positron affinities, on other fit results.
The impact of using positron affinities, for example, can be seen by comparing the crossing point of the Cu (green) and Si (red) fraction curves in all three fraction plots.
From top to center, the higher positron affinity of Si shifts the crossing point towards lower implantation energies, i.e., increases the fraction of positrons annihilating in Si.
Equal affinity values for Cu and Si (bottom) shift the crossing point towards higher energies (even further than for the implantation profile).
%The non-cumulative visualization can be used to determine the implantation energy at which the most positrons are implanted in a layer of interest.

\section{Conclusion}
We have presented \ac{limpid}, a versatile computational tool for modeling positron diffusion and annihilation profiles in layered materials.
By integrating well-established implantation models and a Markov chain approach to diffusion, \ac{limpid} provides accurate fits to depth-resolved \ac{dbs} data.
Future developments will, amongst others, focus on extending support for anisotropic materials (mainly defect distributions), offering the possibility for a second surface (i.e., finite thickness samples), and adding more models for positron implantation.

% Each of the commands below will create an unnumbered section with the appropriate heading.
% Remove any sections that are not relevant for your article.
% All sections except suppdata will be removed if the [anonymous] option is used.
% See iopjournal-guidelines.pdf for more information.

%\ack{Sample text inserted for demonstration.}

%\funding{Sample text inserted for demonstration.}
\section*{Acknowledgment}
Financial support by the German Research Foundation (DFG) within the Project HU 978/19-1 is gratefully acknowledged.

%\roles{Sample text inserted for demonstration.}
% List author names and the contributions made to the article, using terms from the NISO Contributor Roles Taxonomy (CRediT) https://credit.niso.org

%\data{Sample text inserted for demonstration.}
% For more information on IOP Publishing's research data policy see: https://publishingsupport.iopscience.iop.org/questions/research-data/

%\suppdata{Sample text inserted for demonstration.}

%\section*{References}

\printbibliography{}

\end{document}